\documentclass[10pt,letterpaper]{article}
\usepackage[top=0.85in,left=2.75in,footskip=0.75in,marginparwidth=2in]{geometry}

\usepackage[utf8]{inputenc}

\usepackage{cite}

\usepackage{nameref,hyperref}

\usepackage[right]{lineno}

\usepackage[table]{xcolor}

\usepackage{array}

\usepackage{changepage}

\usepackage{textcomp,marvosym}

\usepackage{fixltx2e}

\usepackage{amsmath,amssymb}

\usepackage{cite}

\usepackage{nameref,hyperref}

\usepackage[right]{lineno}

\usepackage{microtype}
\DisableLigatures[f]{encoding = *, family = * }

\usepackage{amsmath}

\usepackage{setspace}

\usepackage{microtype}
\DisableLigatures[f]{encoding = *, family = * }

\raggedright
\usepackage{changepage}

\usepackage[aboveskip=1pt,labelfont=bf,labelsep=period,singlelinecheck=off]{caption}

\makeatletter
\renewcommand{\@biblabel}[1]{\quad#1.}
\makeatother

\usepackage{lastpage,fancyhdr,graphicx}
\usepackage{epstopdf}
\fancyheadoffset[L]{2.25in}
\fancyfootoffset[L]{2.25in}

\usepackage{color}

\definecolor{Gray}{gray}{.25}

\usepackage{graphicx}

\usepackage{sidecap}

\usepackage{wrapfig}
\usepackage[pscoord]{eso-pic}
\usepackage[fulladjust]{marginnote}
\reversemarginpar

\newcommand*{\na}{\ensuremath{\operatorname{NA}}}

\newcommand*{\TCC}{\ensuremath{\mathrm{TCC}}}
\newcommand*{\SVD}{\ensuremath{\mathrm{SVD}}}
\newcommand*{\NAC}{\ensuremath{\mathrm{NA}_c}}

\begin{document}
\vspace*{0.35in}

\begin{flushleft}
{\Large
\textbf\newline{Using Machine-Learning to Optimize phase contrast in a Low-Cost Cellphone Microscope}
}
\newline
Benedict Diederich \textsuperscript{1,2,3,*},
Rolf Wartmann\textsuperscript{1},
Harald Schadwinkel\textsuperscript{1},
Rainer Heintzmann\textsuperscript{2,3},

\bigskip
\textbf{1} Carl Zeiss Microscopy GmbH Research Department, Göttingen, Germany
\\
\textbf{2} Leibniz Institute of Photonic Technology, Albert-Einstein Str. 9, 07745 Jena, Germany
\\
\textbf{3} Institute of Physical Chemistry and Abbe Center of Photonics, Friedrich-Schiller-University Jena, Helmholtzweg 4, 07743 Jena, Germany
\\
\bigskip
* benedict.diederich@ipht-jena.de

\end{flushleft}

\section*{Abstract}
Cellphones equipped with high-quality cameras and powerful CPUs as well as GPUs are widespread. This opens new prospects to use such existing computational and imaging resources to perform medical diagnosis in developing countries at a very low cost. 

Many relevant samples, like biological cells or waterborn parasites, are almost fully transparent. As they do not exhibit  absorption, but alter the light's phase only, they are almost invisible in brightfield microscopy. Expensive equipment and procedures for microscopic contrasting or sample staining often are not available.

Dedicated illumination approaches, tailored to the sample under investigation help to boost the contrast. This is achieved by a programmable illumination source, which also allows to measure the phase gradient using the differential phase contrast (DPC) \cite{Hamilton1984a, Ryde2006a} or even the quantitative phase using the derived qDPC approach \cite{Tian2015}. 

By applying machine-learning techniques, such as a convolutional neural network (CNN), it is possible to learn a relationship between samples to be examined and its optimal light source shapes, in order to increase e.g. phase contrast, from a given dataset to enable real-time applications. 
For the experimental setup, we developed a 3D-printed smartphone microscope for less than 100 \$ using off-the-shelf components only such as a low-cost video projector. The fully automated system assures true Koehler illumination with an LCD as the condenser aperture and a reversed smartphone lens as the microscope objective. We show that the effect of a varied light source shape, using the pre-trained CNN, does not only improve the phase contrast, but also the impression of an improvement in optical resolution without adding any special optics, as demonstrated by measurements.





\section{Introduction}
In recent years the field of smart microscopy tried to enhance the user-friendliness as well as the image quaility of a standard microscope. Since then, the final output of the instrument thus can be more than what the user sees through the eyepiece. Taking a series of images and extract the phase information using the transport of intensity equation (TIE) \cite{Zuo2015a}, extracting the amplitude and phase from a hologram \cite{Liebling2004, Greenbaum2013}  or capture multi-mode images such as darkfield, brightfield and qDPC \cite{Liu2015, Tian2015, Jung2017} at the same time are just some examples. \\ at the same time are just some examples. \\
Translating those concepts to field-portable devices for use in developing countries or education, where cost-effective devices like smartphones are widely spread, brings powerful tools for diagnosis or learning to the masses. \\

Most biological organisms have only low amplitude contrast, thus are hardly  visible in common brightfield configurations. Interferometric or the already mentioned holographic approaches are able to reconstruct the phase in a computational post process. Other methods like the well known differential interference contrast (DIC) or Zernike Phase contrast use special optics to convert the phase into amplitude contrast to visualize these objects. \\

In the year 1899 Siedentopf \cite{Siedentopf1903} suggested a set of rules which can be used to enhance the contrast of an object by manipulating the design of the illumination source. Best contrast for a 2-dimensional sinusoidal grating e.g. can be achieved using a dipole configuration, where two illumination-spots perpendicular to the grating vector are placed in the condenser aperture. Since all objects can be modelled as a sum of an infinite number of sine/cosine patterns, there will always be an optimized source shape, which is, however, not always easy to find by just trial-and-error. \\
In computational lithography this principle is known as source-mask optimization (SMO) which reduces the critical dimension (CD), defined by the smallest feature size of the mask which can be imaged, by optimizing a freeform light source using e.g. a DMD \cite{Garetto2012} with an inverse problem. In lithography the subject of optimization is the mask which is well known, whereas in light microscopy the object, more precisely, its complex transmission function is unknown, which can be estimated using methods mentioned above and described below. \\
We present here an optimization-procedure for the illumination shape. The starting point of the algorithm is given by the estimated complex object transmission function $t(x)$, which is given by inverse filtering of multiple intensity images using the weak-object transfer function (WOTF) developed by Tian et al. \cite{Tian2015}. \\

The paper is organized in four parts, where we start with a short introduction into the theory of image formation in an incoherent optical system and how it depends on the shape of the light source in section two. The third part shows methods for optimizing a light source shape which is then used in two different experimental systems described in part four. The experimental results taken with the lab-microscope which is equipped with a home-made low-cost SLM made by a smartphone Display and the smartphone microscope which derives from a low-cost LED-projector are shown in the fourth and fifth section. \\

\section{Theory}
\subsection{Contrast formation in Partially Coherent Imaging Systems}
\label{sec:theory}

To optimize the contrast of an object at a given illumination configuration, a forward model has to be defined which is used by the optimization routine in order to evaluate the effect of changes in the light source. \\

A common way to simulate a partially coherent imaging system in Koehler configuration, is to compute the system’s transmission cross coefficient (\TCC) introduced by Hopkins \cite{Hopkins1952} and processes the spectrum of the object-transmission function with this 4D transfer function. 
Abbes approach \cite{Abbe1873} discretizes the effective light source into a sum of infinitesimal small point sources $s(\nu_s)$. Each of these point-emitters shifts the object's spectrum in Fourier-space. After summing all intensities of the inverse-Fourier transform with the objecitve's PSF $h(x)$ filtered spectrum over the area of the spatially extended condenser pupil \NAC following \cite{Kumar2010, Hofmann1980}, one gets the partially-coherent image (in lithography often called aerial image).\\
To simplify the derivation of the image formation process, we limit calculus only to frequencies in $x$-direction. The same holds true if one extends the problem to two dimensions. The intensity then computes to
\begin{eqnarray}
I &= \underbrace{\int\limits_{{{\NAC}}} \underbrace{\left|s(\nu_s)\right|^2 }_{\substack{\text{{intensity of condenser}}}} 
    \cdot \underbrace{\left| \int \limits_{-\infty}^\infty t(x') h(x-x') \cdot e^{i2\pi \; \nu' \cdot x'} \; \delta x'\right|^2} _{\substack{\text{{intensity as iFT psf-convolved object-spectra}}}} 
}_{\substack{\text{{summing intensities over all source points}}}}\cdot \delta \nu_s.
\label{eq:mutualcoh3}
\end{eqnarray}
where \NAC$=n\cdot sin(\alpha_c)$ represents the numerical aperture of the condenser. After interchanging the integration boundaries one gets the \textit{TCC} following Hopkins \cite{Hopkins1952} 
\begin{eqnarray}
I(x)  &= \frac{1}{C_n}\int\limits_{-\infty}^\infty  T(\nu_1)T^*(\nu_2)  \left[ \int\limits_{-\infty}^\infty \lvert S(\nu_s)\rvert^2 \cdot H(\nu_1+\nu')H^*(\nu_2+\nu')\cdot \delta \nu'\right] \\
&\cdot e^{i2\pi\; (\nu_1-\nu_2) \cdot x}\;  \delta \nu_1 \delta \nu_2\\
&= \frac{1}{C_n}\int\limits_{-\infty}^\infty  \underbrace{T(\nu_1)T^*(\nu_2)\cdot }_{\text{{bilinear object-spectra}}} \underbrace{ \underbrace{TCC(\nu_1, \nu_2) }_{\text{{system description}}} \cdot e^{i2\pi\; (\nu_1-\nu_2) \cdot x}  \delta \nu_1 \delta \nu_2}_{{{{\cal F}^{-1}}}},
\label{eq:TCC1}
\end{eqnarray}
where the object spectrum is given by a Fourier transform of the tranmission-function $T(\nu)=\mathfrak{{\cal F}}\{t(x)\}$. $H(\nu)$ and $S(\nu)$ represent the two-dimensional pupil functions from the objective and the condenser respectively. Detailed investigation can be found in \cite{Gross2006}.

\begin{figure}[!]
\centering
\includegraphics[width=1\linewidth]{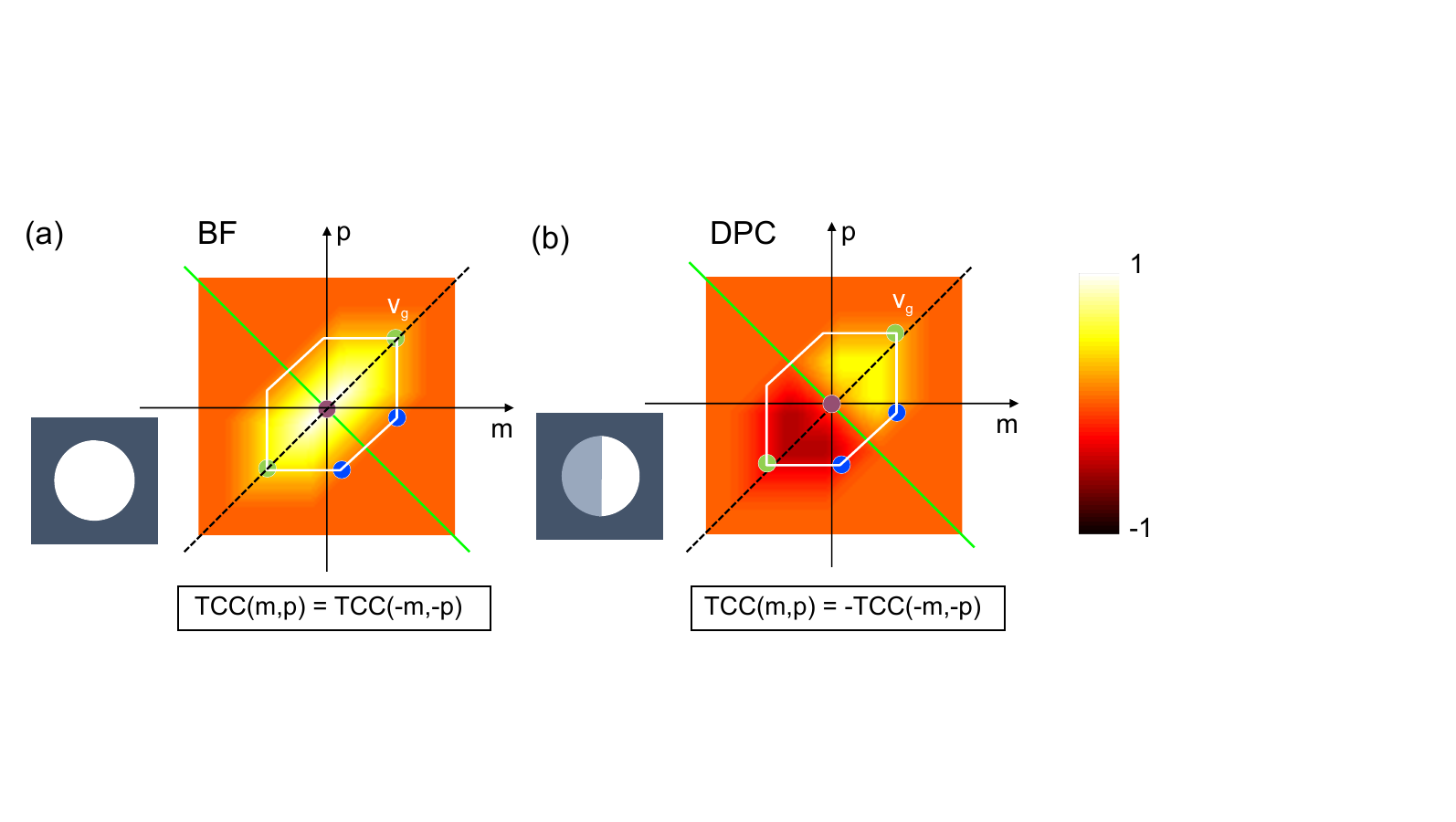}
\caption{{\bf Symmetry properties of the TCC at different illumination configurations} In (a) the \TCC\; at $p=q=0$ gives the partially coherent transfer function for a brightfield and in (b) for a DPC system (b). The green line shows the axis of symmetry. The DPC setup offers odd symmetry which enables phase-contrast.}
\label{fig1}
\end{figure}

Following \cite{Hopkins1952} the so-called Transmission-Cross-Coefficient $TCC(\nu_1, \nu_2)$ or sometimes partial coherent transfer function $PCTF$ \cite{Kumar2010}, can be extracted from Eq.\,\ref{eq:TCC1} which then describes the imaging properties of a partially coherent imaging system and contains all information about illumination and aberration:
\begin{eqnarray}
TCC(\nu_1, \nu_2)= \frac{1}{C_n}\cdot \int \lvert S(\nu_s)\rvert^2 \cdot H(\nu_1+\nu')H^*(\nu_2+\nu')\; \delta \nu',
\label{eq:TCC2}
\end{eqnarray}
where $C_n$ is the normalization-factor, that is usually the maximum value of the transfer function \cite{Gross2006}.\\ 

This filter function can be understood as the geometrical overlap of the 2D objective pupil $H$, its conjugate $H^*$ and the effective illumination source $S$. The shift of the $H$ and $H^*$ corresponds to a spatial frequency caused by the bilinear object spectrum. Thus the $\TCC(\nu_1, \nu_2)$ gives the attenuation of an object frequency pair at a certain spatial frequency.\\
The main advantage of this method is, that the \TCC, once computed, can be stored in memory for later reuse, thus it represents a very computationally efficient way to see how the contrast is modulated by the microscope. \\
A simplification introduced by Cobb et al. \cite{Cobb1998} and later further developed by Yamazoe et al. \cite{Yamazoe2008a, Yamazoe2009} reduces this 4D matrix into a set of 2D-filter kernels by applying the singular-value decomposition (SVD). The result will be a set of eigenvectors which acts as filter kernels and its eigenvalues which gives the weights. It was shown, that the eigenvalues decrease rapidly, thus it’s possible to use only the first 2 Eigenvectors to simulate an image by keeping the error below 10 \%. 

\subsection{Partially Coherent Image Formation}
\label{sec:optimization}
\begin{figure}[!]
\centering
\includegraphics[width=0.95\linewidth]{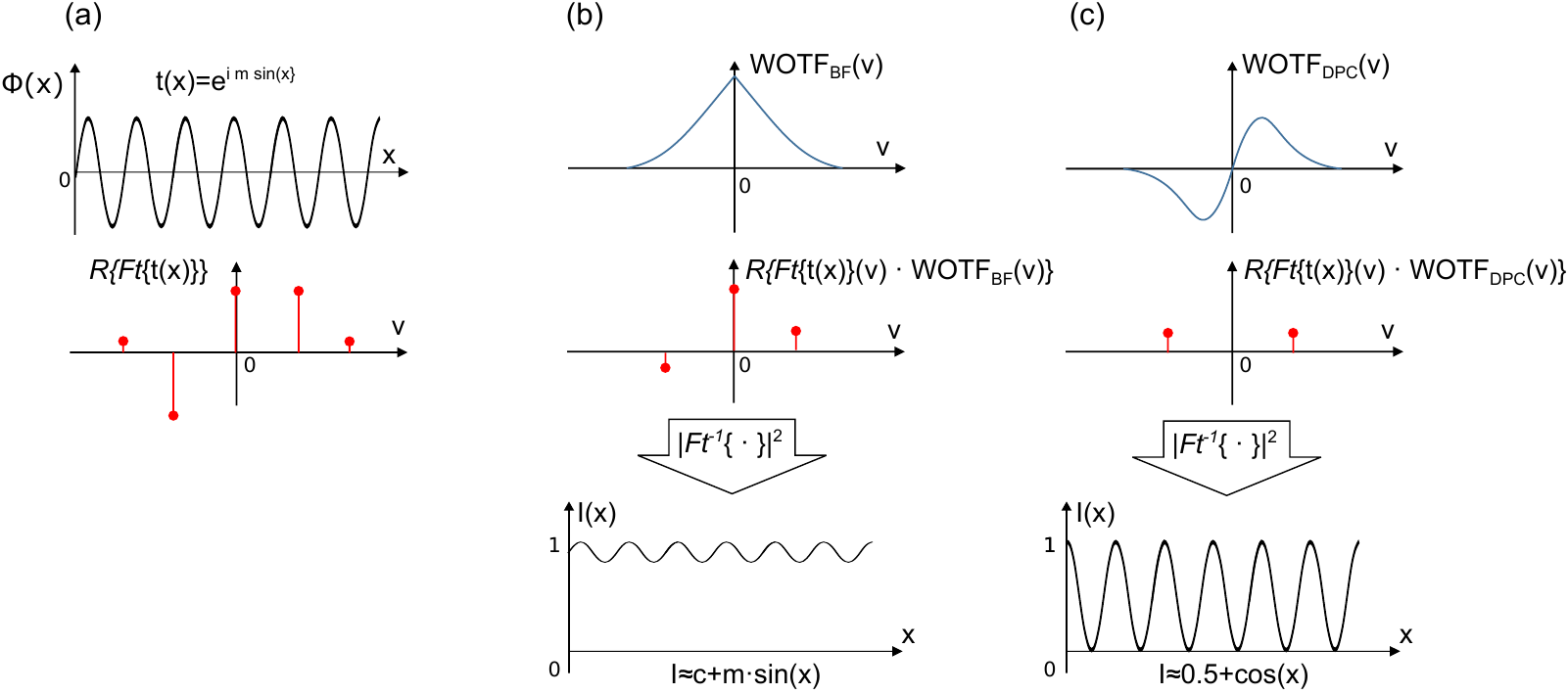}
\caption{\textbf{Asymmetric illumination source enables phase-contrast.} (a) shows the transmission-function $t(x)$ of the sinusoidial phase object and its spectrum which gets filtered in (b) by the WOTF of the brightfield microscope  and in (c) by the DPC-system. One clearly sees, that an odd symmetric optical system is capable of transmitting phase information and images the phase-gradient.}
\label{fig2}
\end{figure}

It can be shown, that without adding any additional optics, such as phase masks in Zernike phase contrast or prisms in DIC, the object’s phase contrast can be enhanced by an iterative optimization process which manipulates the shape of the condenser aperture. \\
A mathematical motivation of this phenomena can be found in Sheppard/ Wilson \cite{Hamilton1984a} which claims that a symmetric and real-valued optical system, such as the perfect brightfield microscope, is not capable to image the phase of an object. To make phase variations visible one can make the \TCC\; asymmetric by illuminating the object asymmetrically (e.g. DPC) or adding a phase factor in the pupil plane (e.g. introducing a phase mask or defocus the object).\\
This can be proved by a simple example, where we define a sinusoidal phase grating visualized in Fig \ref{fig2} $t(x)=A_0\cdot exp\left(i\frac{m}{2}sin\left(\frac{x}{{k_g}}\right)\right)$. Its direction is defined by the k-vector, which can be represented as a scalar in the one-dimensional case ${k_g}$, $m\subset[0, 1]$ defines its modulation depth . Its Fourier-transform is given by a Taylor series expansion
\begin{eqnarray}
\mathfrak{F}\{t(x)\}(\nu) &= T(\nu_x)=\sum_{-\infty}^{\infty} J_q\left(\frac{m}{2}\right)\cdot \delta\left(\nu_x-q\cdot \frac{1}{{k_g}}\right),
\label{eq:spectragrating}
\end{eqnarray}\\
where $J$ is the Bessel function of the 1-st order. The aerial image (Fig \ref{fig2}) is given by
\begin{eqnarray}
I(\nu_x) &= \int_{-\infty}^{\infty} TCC(\nu_x+\nu_1;\,\nu_2)T(\nu_x + \nu_1) T^*(\,\nu_2)\delta \nu_2.
\label{eq:intensity}
\end{eqnarray}
Taking the $q=0$, $q=\pm1$ diffraction order into account only (indicated as red/blue dots in Fig \ref{fig1}) and inserting Eq. \ref{eq:spectragrating} into Eq. \ref{eq:intensity}, the formula reduces to 2 cases:\\
\begin{eqnarray}
&1.\; \nu_x=0& \\
&I(\nu_x) &= C_1\cdot TCC(0,0) + C_2^2\left(TCC\left(\frac{1}{{k_g}}; \frac{1}{{k_g}}\right)-TCC\left(-\frac{1}{{k_g}}; -\frac{1}{{k_g}}\right)\right)\\
&2.\; \nu_x=\pm \frac{1}{{k_g}}& \\
&I(\nu_x) &= C_1C_2\cdot TCC\left(\frac{1}{{k_g}},0\right) - C_1C_2\cdot TCC\left(0; \frac{1}{{k_g}}\right)
\label{eq:tccsym}
\end{eqnarray}

The \TCC\;of a ``perfect'' brightfield microscope has even symmetry and has a real-valued pupilshape, which means, that no phase factors e.g. aberrations are present. Inserting $TCC(m,p) = TCC(-m,-p)$ into the two cases show, that the phase term vanishes and only a DC-term predominates the aerial image.\\
By sequentially shading both sides of the condenser aperture and subtracting the resulting two images from one another gives an odd-symmetric \TCC\;as can be seen in the 2D-\TCC\; plot in Fig \ref{fig1} (b). Inserting $TCC(m, p) = -TCC(-m, -p)$\; into eq. \ref{eq:tccsym}, one can see, that the intensity follows the phase gradient plus a DC-term $(I_1+I_2)$ which can also be removed \cite{Mehta2009b}
\begin{eqnarray}
I_{DPC} = (I_1- I_2)/(I_1+I_2).
\label{eq:DPC}
\end{eqnarray}
This brief mathematical description shows, that an appropriate light source shape enables to image phase contrast by manipulating the condenser's aperture plane which can conveniently be done with an SLM, also shown in \cite{Bian2015} and \cite{Iglesias2013}.

\section{Optimization of the light source}
\label{sec:optalgo}
According to the illumination principle introduced by Siedentopf \cite{Behrens1915} it is best practise in order to enhance the contrast in the object plane, by leaving out the illumination direction which do not contribute to the image formation.\\

This gives a huge degree of freedom which can be reduced by simplifying the effective area by a finite number of parameters which represent a pattern, when trying to optimize a $m\times n$ pixel grid of a freeform light source. In \cite{Wu2014} this has been done by by a set of Zernike Coefficients which reduces the number of parameters considerably. An alternative is to divide the circular condenser aperture in ring segments and weigh each of them with a specific value. 

A generalized light source can be described as a superposition of individual sub-illumination functions $Z_s(m,n)$, where each can be weighted with a factor $\Psi_l$:
\begin{eqnarray}
Z_s(m,n) = \sum_{l=1}^P\Psi_l\cdot S_l = S\cdot \Psi,
\label{eq:source}
\end{eqnarray}
and the latter one represents the source in matrix notation with $\Psi = \left[\Psi_1\, \Psi_2\, ...\, \Psi_P\right]^T$ and $m,\,n,\,p,\,q\,$ give the discrete pixel coordinates. $S$ holds the light patterns, e.g. i-th Zernike polynomial or circular segment.

Inserting eq. \ref{eq:source} in eq. \ref{eq:TCC2} gives a \TCC\;whose imaging properties now depend on the parameter-set $\Psi$ multiplied with the precomputed four-dimensional $TCC_S(m,n;p,q)$, which allows the computation of two-dimensional images
\begin{eqnarray}
TCC_{sys}(m,n;p,q) =& \frac{1}{C_n}\sum_{m,n;p,q} Z_S^2(m,n;p,q)\cdot H(m+m',\,n+ n') \cdot H^*(p+p',\,q+q')\\
=&\underbrace{TCC_{s,i}}_{\substack{\text{precalculated TCC}}}\cdot \underbrace{\Psi}_{\substack{\text{vector with P-coefficients}}}
\label{eq:zernikeopt3}
\end{eqnarray}
Separating the $\TCC_Z$ into its Eigenvalues $\lambda_i$ and Eigenfunctions $\Psi_i$ using the \SVD\; gives
\begin{eqnarray}
TCC_{\SVD}(m,n;p,q) &= \sum_{j}^{P}\sum_{i=1}^{k}\lambda_{i}\{\Psi_j\}\Phi_{i}(m,n)\{\Psi_n\}\Phi_{i}^*(p,q)\{\Psi_j\},
\label{eq:svd}
\end{eqnarray}
thus the intensity of the aerial image reduces to 
\begin{eqnarray}
I_{sim}(x',y')=& \sum_{j}^{P}\sum_{i=1}^{k}\lambda_{i}\{\Psi_j\}\left|\mathcal{F}\left\{T(m,n)\cdot\Phi_{i}(m,n)\{\Psi_j\}\right\}\right|^2\\
=&\sum_{j}^{P}\sum_{i=1}^{k}\lambda_{i}\{\Psi_j\}\left|t(x,y)\otimes\phi_{i}(m,n)\{\Psi_j\}\right|^2.
\label{eq:zernikeopt5}
\end{eqnarray}
where $k$ is the maximum number of used filter kernels (e.g. \SVD-Eigenvectors).

Varying the source parameters has an influence on contrast, but also on intensity in the specimen plane or spatial resolution, where each image's quality criterion can be measured by a tailored cost function. One example for a possible cost function can be derived following the so-called fidelity \cite{Gross2006}, which is defined as
\begin{flalign}
F &= ||I-I_{ideal}||^2  \\
I_{ideal}(x) &= |t(x)|^2 \text{ bzw. } I_{ideal}(x) = arg(t(x)) \text{ when using phase objects} \\
F&=\sqrt{\sum_x\,I(x)-|t(x)|^2}.
\label{eq:fidelityqua}
\end{flalign}
This gives a quality measure how similar the intensity measurement, e.g. on the camera, follows the quantitative phase estimate by the qDPC image. Another approach could be to maximize the standard deviation of the simulated pixels and thus maximize the phase contrast, by changing the pixels inside the aperture plane. The next step would be to create a cost function using eq. \ref{eq:fidelityqua} and eq. \ref{eq:zernikeopt5}. \\ 
\begin{flalign}
F(\Psi) &= ||I_{sim}\{\Psi\}-I_{ideal}||^2  \\
&= \sum_{i=1}^{x,y} (I_{sim}\{\Psi\}-I_{ideal})^2  \\
\intertext{using Eq. \ref{eq:zernikeopt5}}
&= \sum_{i=1}^{x,y}\left(\sum_{n}^{P}\sum_{i=1}^{k}\underbrace{\lambda_{i}}_{{\substack{\text{\SVD}}}} \{\Psi\}\left|t(x,y)\otimes\underbrace{\phi_{i}}_{{\substack{\text{\SVD}}}}(m,n)\{\Psi_n\}\right|^2-I_{ideal}\right)^2  \\
&= \sum_{i=1}^{x,y}\left(\sum_{n}^{P}\sum_{i=1}^{k}\SVD\left[\TCC(\Psi_n)\right]_{\lambda_{i}} \left|t(x,y)\otimes \SVD\left[\TCC(\Psi_n)\right]_{\phi_{i}})(m,n)\right|^2-I_{ideal}\right)^2.
\label{eq:fidelity}
\end{flalign}

\subsubsection{Particle Swarm Algorithm Optimization}
The simplification in the previous step, to accelerate the computation of an aerial image by using the \SVD, complicates the calculation of the analytical gradient $\nabla\{F(\Psi)\}$ to feed e.g. a gradient-descent algorithm. Using genetic algorithms, like the so-called particle swarm optimization algorithm (PSO) \cite{Eberhart2007}, bypasses the lack of the analytical gradient.\\ 

The PSO finds its analogy in the behaviour of a swarm, such as a flock of birds, where the direction and velocity of its centre depends on the movement of each single individual as well as the whole swarm. In our approach, each individual of the swarm evaluates one possible illumination situation using the parameter-vector $\Psi$. The swarm size was empirically chosen to be $15\cdot l_{max}$, where $l_{max}$ corresponds to the maximum numbers of variable parameters, as suggested by Montgommery \cite{Chen2015}.\\
The iteration was stopped, once the movement of the swarm slows down to a specific amount or by reaching a maximum number of iterations \cite{Eberhart2007}. The algorithm developed in Matlab 2015a (The MathWorks, Inc., Massachusetts, USA) and Tensorflow \cite{Abadi} is available on Github \cite{Diederich2017, Diederich2017a}.

\subsubsection{Gradient Descent Optimization}
\label{sec:gdo}
Alternatively we used the auto-differentiation functionality of the open-source ML library Tensorflow \cite{Abadi} which enables a gradient based optimization using different optimizers (i.e. Gradient-Descent, Adadelta, etc.). We used the precomputed eigenfunctions and eigenvalues from eq. \ref{eq:svd} to predict the intensity of the aerial image in eq. \ref{eq:zernikeopt5} following eq. \ref{eq:zernikeopt5}. Different error-norms can be introduced, while the so called uniform-norm, which tries to maximize the absolute difference between the minimum and maximum Intensity value of the simulated image\\
\begin{flalign}
F(\Psi) &= -||max(I_{sim}\{\Psi\})-min(I_{sim}\{\Psi\})||,  \\
\end{flalign}
was chosen as the best error measure for the given situation. In both cases, the initial values of the illumination source were set to one.

%
%

\section{Using Machine Learning to Optimize the light source}
\label{sec:machinelearn}

Even after optimizing the code by replacing the 4D-\TCC\;with only two convolution-kernels following eq. \ref{eq:svd} to simulate one intensity image, the process of finding an optimized set of the parameters takes about $20\,s$ on a quadcore computer, which is not reasonable for biologists or high throughput-applications, such as drug screening or industrial metrology, who needs results in real-time.\\
Recent years gave rise to computational power and brought the field of Big Data in combination with neural networks back to focus. The field of machine learning has been applied to many tasks in microscopy such as automated cell-labelling \cite{Sommer2013a} or learning approaches in optical tomography \cite{Kamilov2015}. In our approach we show, that a neural network is capable to learn an optimized light source shape from a set of prior generated input-output-pairs. A major advantage of a neural network is, once it is trained, the learned weights can be used on devices with low computational power, such as cellphones. \\

\subsection{Generating a Dataset for Machine Learning}
\label{sec:genmachinelearn}
For training the network, we generated a data-set of microscopic-like objects. We exploit the self-similarity of biological objects \cite{Losa2011}, especially in its  ultrastructure and overall shape, to maximize the generality of our application by building a universal dataset. This is possible, because an optimized illumination source rather depends on the object's structure, than its type. \\

Therefore we included about 1.000 randomly chosen microscopy images (i.e. epithelial cells, bacteria, diatoms, etc.) from public databases \cite{BrandnerD.andWithers2010, LarryPage} as well as from acquired images using the qDPC approach. To expand the dataset, we added artificially created patterns from a 2D discrete cosine transformation (DCT) which mimic grating-like structures, such as muscle-cells or metallurgical objects, very well. \\

The datasets also includes the optimized illumination-parameters to enhance the phase-contrast of the complex object transfer function, generated with the algorithms explained in section \ref{sec:optimization}. The transfer function itself was derived by generating pseudo phase maps from the intensity images.\\
Augmentation of the samples to avoid overfitting in the learning process of the NN was done by adding noise and rotation to the images. In case of the segment-illumination, the parameters have to be rotated likewise, due to the Fourier-relationship, which states, that a rotation around the zero-position corresponds to a rotation in frequency-space, thus no reevaluation of the parameters using the algorithm is necessary. \\

To overcome possible phase ambiguities, the complex object was split into real and imaginary part rather than amplitude and phase, which then creates a 2-layer 2D image and can be converted into a $N\times N\times 2$ matrix, where $N$ corresponds to the number of pixels along $x$ or $y$. \\

\subsection{Architecture and Training of the CNN}
\label{sec:cnnarchitecture}

The cost function was defined as the mean-squared error (MSE) of the predicted  $\hat m$ and measured values $m$ of the output intensities in the condenser aperture plane as 
\begin{flalign}
\operatorname{MSE}=\sum_{i=1}^n (\hat m - m)^2. 
\label{eq:rms}
\end{flalign}
It turned out, that the criteria to predict the CNN's accuracy, by simply comparing the exact matches of the continuous output and the dataset values would be too strict. Therefore the continuous output in the range of $\hat y \subset [0..1]$ was quantized by rounding the first decimal place. Thus the accuracy is calculated by summing over all equal output-pairs and then normalizing it by dividing by the number of samples to get the relative accuracy
\begin{flalign}
\operatorname{ACC}=\frac{1}{n}\sum_{i=1}^n \left(\operatorname{round}(\hat m_i \cdot 10) = \operatorname{round}(m \cdot 10)\right). 
\label{eq:acc}
\end{flalign}

The dataset was separated into a train-, validation- and test-set by the fractions 80/10/10\,\%. A batch-size of 128 training-samples was feed into the CNN, where the error was optimized using the ADAM \cite{Kingma2014} update rule with a empirically chosen learning rate of $10^{-2}$. The architecture was derived from the standard fully supervised convolutional neural network (convnet) by LeCun et al. \cite{LeCunJackelB.BoserJ.S.DenkerD.HendersonR.E.HowardW.Hubbard1990}) and Krizhevsky et al \cite{Krizhevsky2012} and implemented in the open-source software Tensorflow \cite{Abadi} which works also on mobile devices with Android or iOS. \\

\begin{figure}[!htb]
\centering
\includegraphics[width=0.95\linewidth]{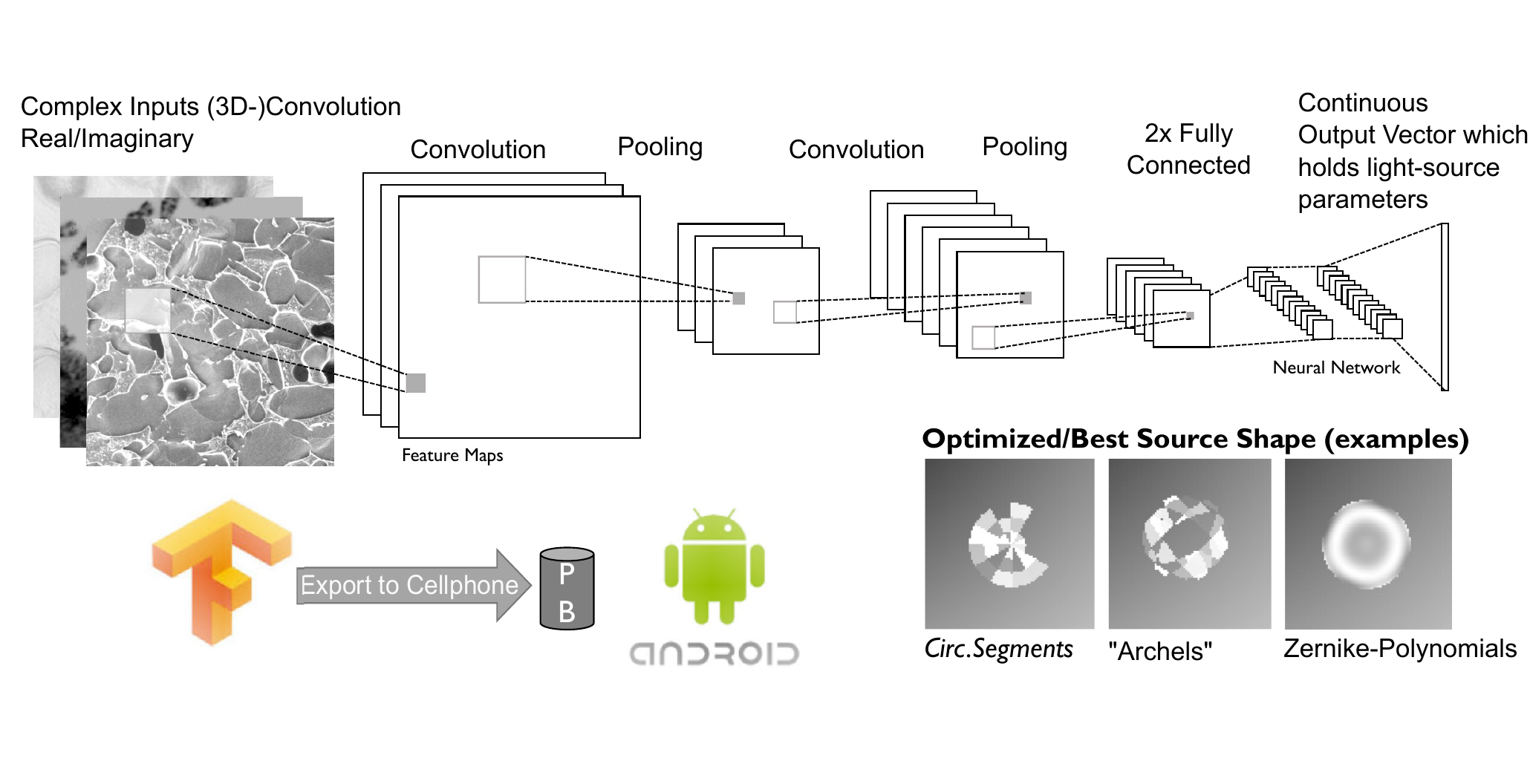}
\caption{\textbf{Basic architecture of the used CNN} CNN which takes the complex 2-channel input images and the generated optimized light source parameters as the training data. The learned filters can then be exported to mobile devices i.e. Android smartphones.}
\label{fig3}
\end{figure}

The complex 2D input image (two channels) $x_i$ is followed by a series of two 2D-conv-layers including a maxpool-layer with a feature size of 128 and kernel size of $9\times9$ and $7\times7$ respectively as shown in Fig. \ref{fig3}. As activation function the $tanh$-function was chosen as this, compared e.g. to the linear $relu$-activation function, is better suited for continuous output values suggested in \cite{Krizhevsky2012}. The top few layers of the network are conventional fully-connected networks with using dropout of 0.5 probability while training.

The calculation was carried out on a Nvidia GTX TitanX, where the relative accuracy converged after about 10k iterations to a value of 67 \% in the testing phase, where we used the discretized error-norm described earlier.\\
The output of the training procedure is a set of filters, where each corresponds to object features which can be enhanced by using an appropriate source shape. \\

As mentioned earlier, there are always many possible solutions that could increase the phase contrast of an object, thus the output describes only a certain probability of an optimized parameter-set $\Psi$. Therefore we calculated the optimized illumination source for the 2D complex images from the test-set using the iterative algorithm in Sec \ref{sec:gdo}. We compared it to the intensity simulations acquired with the illumination source generated using the trained CNN. The learned source-parameters even improved the overall contrast, defined as the difference of the min/max pixel-value improvement of the cost-function, in this case the uniform norm average, was about 2.5 \%. \\

The optimization procedure now reduces to a simple convolution, multiplication and summation of the input-vectors with the learned weights and the application of discriminator functions. Thus it can easily be performed by a mobile device such as a cellphone. The evaluation of the new output values resulting from the learned weights takes now $t_{comp}=30\,ms$ on a computer and $t_{comp}\approx430\,ms$ on a cellphone, whereas the original algorithm in section \ref{sec:optalgo} took more than $t_{comp}=20\,s$.


\section{Methods}

For proving, that the numerical optimization enhances the phase-contrast of a brightfield microscope, a standard upright research microscope ZEISS Axio Examiner.Z1 equiped with home-made SLM using a high-resolution smartphone display (iPhone 4S, Apple, USA) in the condenser plane, was used for the first tests. In comparison to LED-condensers, as presented in \cite{Phillips2015}, a ZEMAX simulation of the koehler-illumination using an LCD in the aperture plane shows two-times better light-efficiency, which also improves the SNR and allows lower exposure times of the camera.\\
The higher pixel-density compared to the LED-matrix also allows the calculation using the \TCC\; without introducing large errors or artefacts in the acquired intensity images and avoids artifacts due to missing sampling points.\\

\subsection{Setup of a Portable Smartphone Microscope}

The effect of using the trained neural network shows its full potential, when it comes to devices with low computational power such as modern cellphones. The Tensorflow library \cite{Abadi} allows to train the network on a desktop-machine and exploit the trained weights/filters on mobile hardware. Low Cost smartphone microscopes were shown in \cite{Phillips2015, Breslauer2009} with single LED as well as LED-matrix illumination and in a holographic on-chip configuration \cite{Tseng2010}.\\
Here the Koehler illumination of a standard microscope is adapted to the mobile device, where a low-cost LED video projector (Jiam UC28+, China) was slightly modified by adding two additional lenses. To limit the spectral bandwidth of the illumination, necessary for getting the estimated phase from the DPC measurement, a green color filter (LEE HT 026,$ \lambda=530\,nm\,\pm20\,nm$) was chosen to keep the system's efficiency as high as possible. 
\\

\subsubsection{Optical and Mechanical Design}
Following the approach by using a reversed smartphone lens ($f\#=k_{eff}=2.0$) of an iPhone 6 (Apple, USA) in \cite{Switz2014} as the microscope objective, one can ensure a diffraction limited -1:1 imaging at a numerical aperture of $\na_{cellphone}=\frac{1}{2k_{eff}}=\frac{1}{2*2.0}=0.25$. \\

The pattern of the projector which itself has an LC-Display (ILI9341), with a transmission of about 3\,\% at 530\,nm and a pixel size of $16\,\mu m$, illuminated by a high-power white light LED equipped with a ground glass diffuser can easily be controlled by the HDMI-port (MPI) of the smartphone. To guarantee a time-synchronized capturing while manipulating the pattern, a customized App \cite{Diederich2017b} was written to trigger the camera-frame with the HDMI-signal using Google's Presentation API. \\

\begin{figure}[!htb]
\centering
\includegraphics[width=0.9\linewidth]{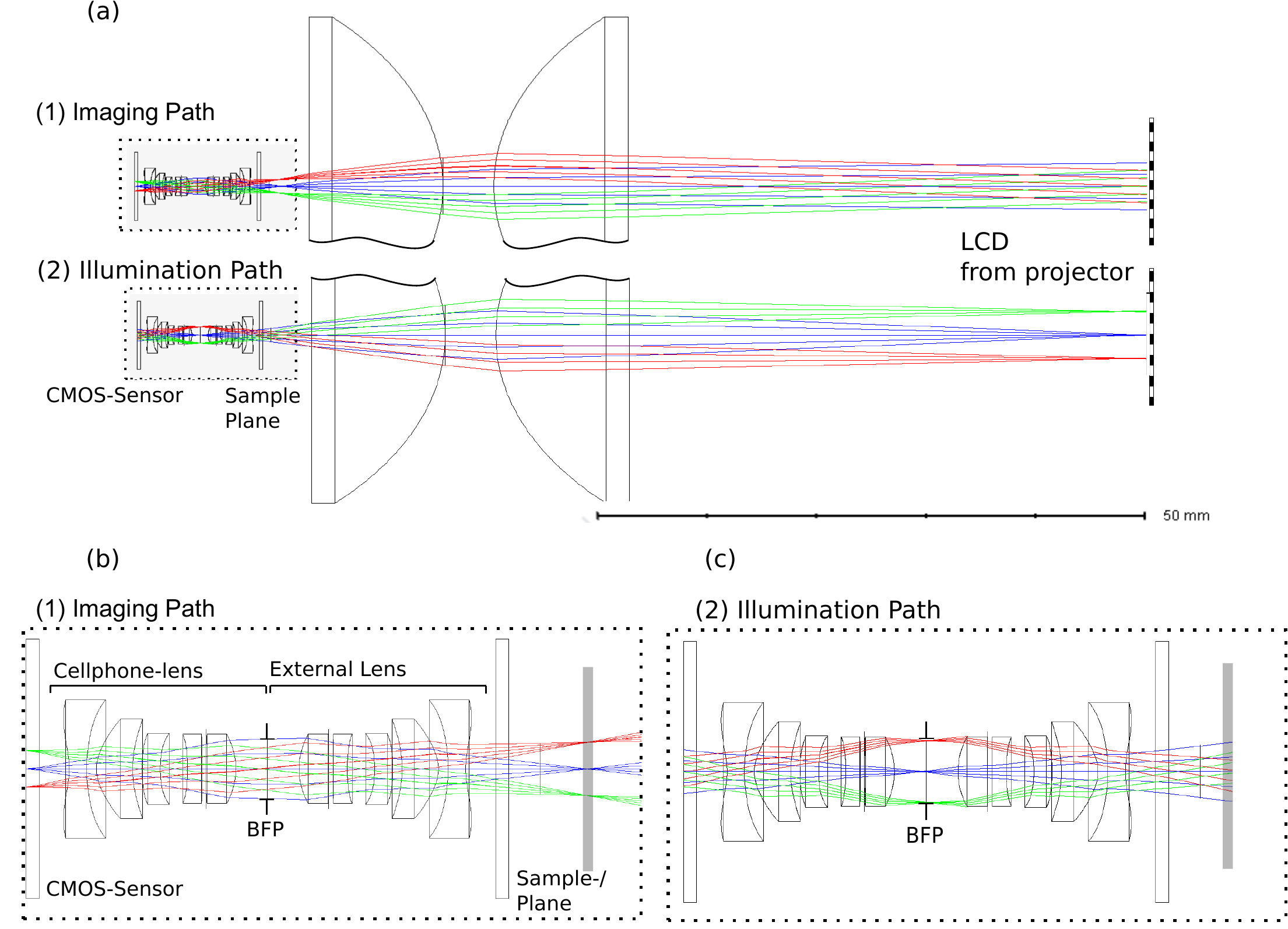}
\caption{\textbf{Optical Design of the illumination system} (a) shows the microscope setup using an inversed camera-lens, the location of the sample-/slide- plane and the LCD from the video-projector. (b) shows the enlarged imaging path from (a), where the objects gets imaged onto the sensor. (c) shows, that the illumination system images the condenser aperture (e.g. the LCD) into the BFP of the microscope objective.}
\label{fig4}
\end{figure}

The housing of the video-projector does not allow an appropriate arrangement of the lenses to form even illumination in the sample plane which is mandatory to realize Koehler Illumination. Therefore a new optical design was developed with commercially available lenses.\\ 
To ensure Koehler condition the LCD is imaged into the entrance pupil of the cellphone-camera which falls into the back focal plane (BFP) of the reversed cell-phone lens. At the same time, the field diaphragm, which confines the visible field of view and reduces stray light, has to be imaged to infinity. Unlike most conventional microscope schemes, the optical setup is not telecentric and the cellphone-lens has a large field-angle of about $70^\circ$. The developed optical design in Fig \ref{fig4} composed by two injection molded aspherical (Thorlabs ACL3026U) and the two cellphone lenses (microscope-objective: iPhone 6 lens, tube lens: LG Nexus 5 lens) satisfies these two requirements. It has to be said, that even though the active area of the LCD which forms the effective light source uses only a fraction of the LCD, thus only few pixels contribute to the image formation. \\ 

\begin{figure}[!thb]
\centering
\includegraphics[width=0.95\linewidth]{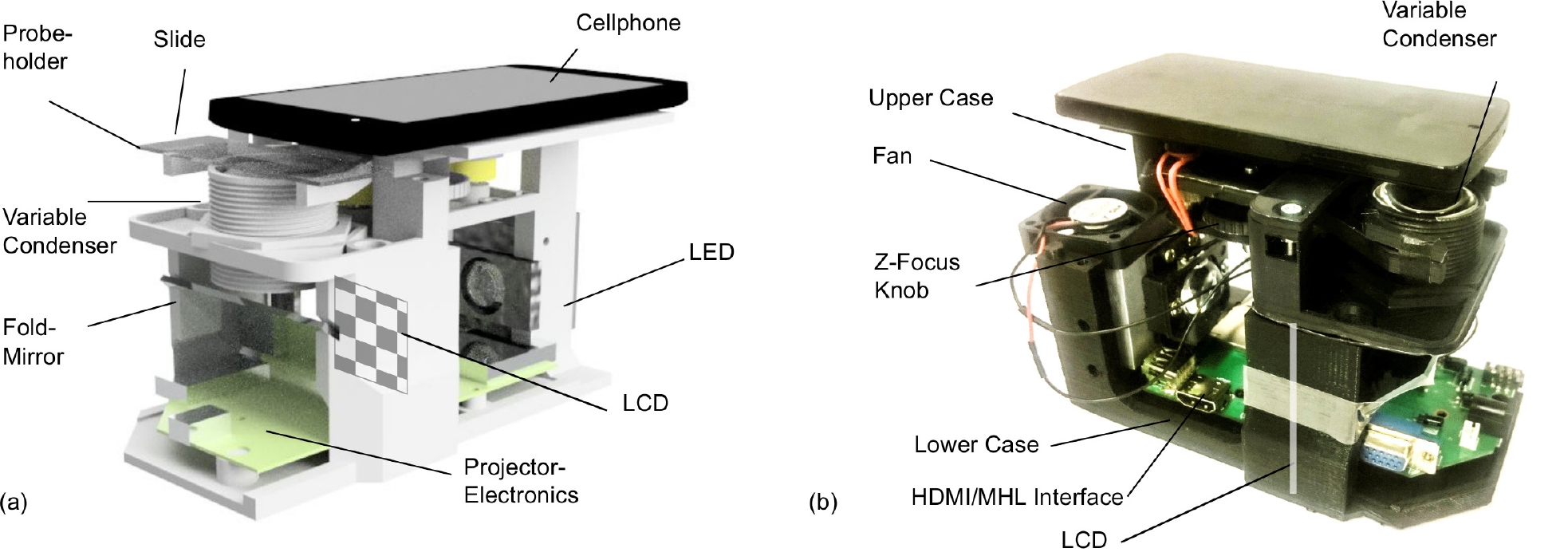}
\caption{\textbf{Rendering and 3D printed model of the microscope} In (a) CAD rendering of the microscope, where the lens-distances were exported from the ZEMAX raytracing software to assure correct optical relationships. In (b) the fully automated microscope which uses a low-cost projector to quantitatively image the object's phase. The location of the LCD is visualized as a white chessboard before a fold-mirror couples the light into the condenser.}
\label{fig5}
\end{figure}

An open-source 3D-printed case (Fig \ref{fig5}, available at \cite{Diederich}) was built to maintain the relationship between all optical components as well as to make the microscope portable. To adjust the Koehler condition the housing of the condenser has a 3D printed outer threading, which allows to change the z-position by simply rotating the lenses. Therefore it is possible to setup the optical distances even if the cellphone is not in the correct position or another phone is used. To ensure optimal centring of the optical axis, the reversed cellphone lens was brought into position with a customized magnetic snap-fit adapter.\\
The focussing of the object was also migrated by a 3D-printed linear stage which uses dry linear bearing (IGUS Drylin, Germany) on polished chrome-rods in order to achieve a precise linear moving. The adjustment is ensured by a micrometer screw (Mxfans, Micrometer 0-25mm, China).\\

\subsubsection{Electrical and Software Design} 
The camera used in our experiments is provided by a LG D850/Google Nexus 5 \cite{Electronics} which is, by now, the only smartphone camera on the market which fully supports Androids Camera2 API, thus allows the readout of the raw pixel values of the sensor before post-processing e.g. demosaicing \cite{Skandarajah2014a}.\\
Camera-control settings like the ISO-value and exposure time, as well as the focus position can be manipulated manually, which gives rise to ensure a linear relationship between the illumination pattern and gray value necessary to satisfy the model requirements in \cite{Skandarajah2014a}. A gamma-calibration is performed, by taking an image sequence of 256 gray patterns displayed on the projector, and estimating the look up table (LUT) by fitting the pixel average of each image to a Sigmoid-function, which characterizes the display of the LCD appropriately. 

From the acquired images, an algorithm adapted from Tian et al. calculates an estimate of the quantiative phase using the OpenCV4Android framework \cite{3.0} and then feeds the pretrained neural network. The final result is a parameter vector which represents the intensity weights of the selected illumination pattern (e.g. circular segments).

The optimized result is then displayed on the LCD (e.g. the condenser aperture) using the MPI interface. Thus the entire microscope can work completely autonomous.

\section{Results and Discussion}

The proposed methods to improve the phase contrast by manipulating the condenser aperture was first tested on a standard research microscope (ZEISS Axio Examiner.Z1) with a magnification $M=20\times,\, \na=0.5$ in air (ZEISS Plan-Neofluar) at a center-wavelength of $\lambda=530nm\pm 20nm$ to give a proof-of-principle before the method was evaluated on the cellphone microscope.\\

The relatively low transmission and extinction, which is usually defined as the ratio of pixel on-/off-state $t_{ex}=\frac{I_{on}}{I_{off}}$, coming from the iPhone ($t_{trans}=6\,\%$, $t_{ex}=1:800$ at $\lambda_0=530nm$) and the projector ($t_{trans}=3\,\%$, $t_{ex}=1:250$ at $\lambda_0=530nm$) result in a low signal-to-noise ratio (SNR) and also in longer exposure times (e.g. $>10\times$ higher) compared to a mechanical condenser aperture. This effect was compensated by the high-power LED in the low-cost setup, where it was possible to get exposure times below $t_{exp}<10\,ms$.\\

An LCD can only approximate a continuous aperture shape such as a circular aperture, but in contrast to the LED-illumination shown in \cite{Phillips2015, Tian2015}, the pixel density is still higher and results in a better approximation of the fundamental imaging theory to calculate e.g. the quantitative DPC.


\subsection{Improvement in Phase Contrast on the Cellphone Microscope}
The method presented here allows for increasing the phase contrast by varying the illumination source and has some major advantages compared to computational post-processing, where phase contrast is enhanced. The SNR of the acquired images can always be maximized in the acquisition process by adjusting the exposure time and per-pixel gain, once the best light source is found, which also results in the optimal dynamic range of the camera sensor. \\
Further advantages of the incoherent light source are, that no speckle patterns in the image are visible and that it gives two times the optical resolution compared to coherent imaging methods \cite{Tian2015a}.\\

\subsubsection{Quantitative Results of Microscopic Objects}
\begin{figure}[!htb]
\centering
\includegraphics[width=1\linewidth]{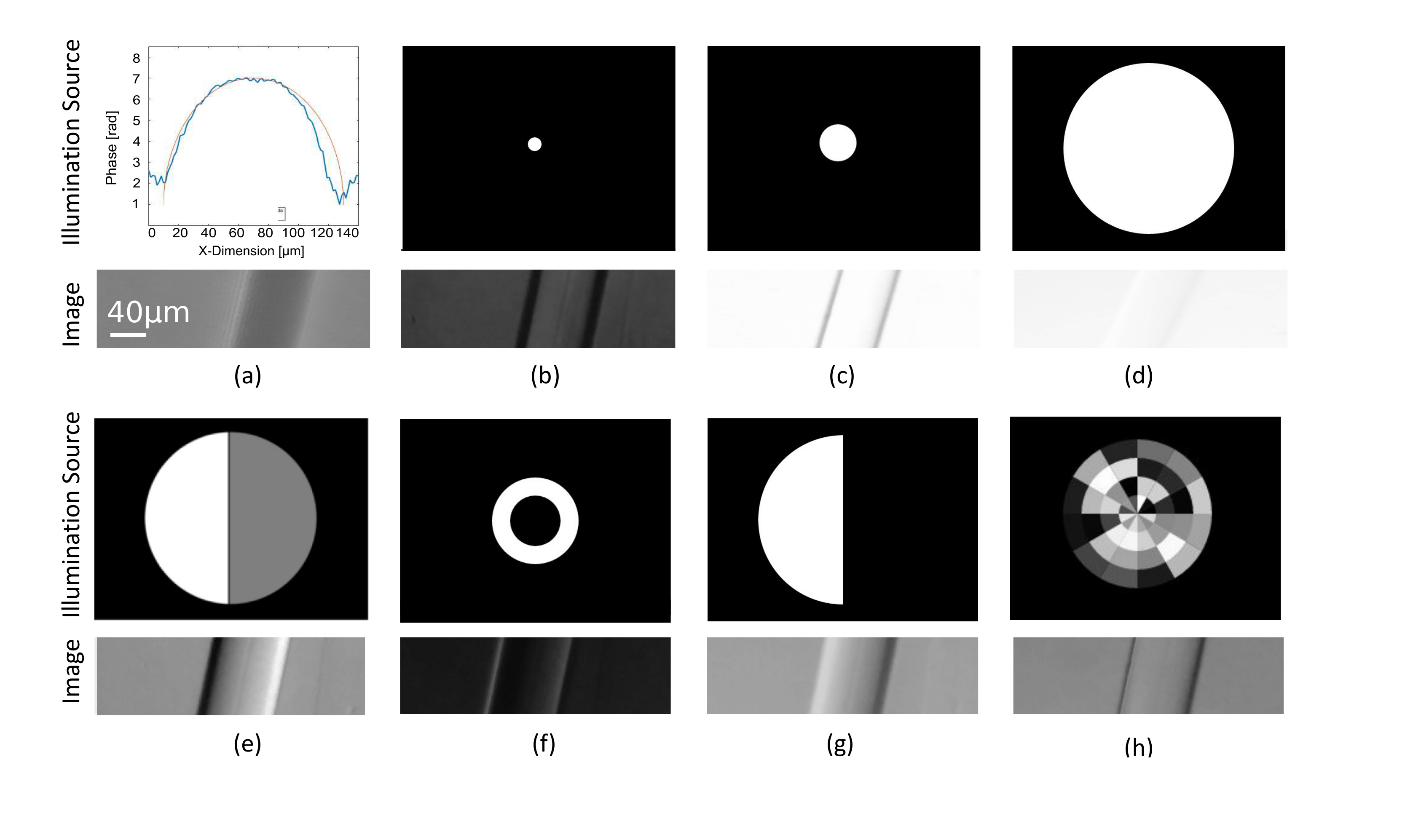}
\caption{\textbf{Quantitative and qualitative results produced by the portable microscope} Quantitatively measured phase of a glass fiber immersed in oil using qDPC mode in (a); Intensity measurements and their corresponding illumination sources using brightfield mode with $\na_C=0.1$ (b), $\na_C=0.2$ (c), $\na_C=0.5$ (d); The computed DPC image in (e), a measurement in Darkfield mode ($\na_o<\na_C$) in (f), oblique illumination in (g) and the optimized light-source ($\na_C=0.3$; magnified for better visualization) using the CNN in (h) using (a) as the input image.}
\label{fig6}
\end{figure}

In order to improve the phase contrast of an object, both, the iterative and machine-learning based version of our algorithm, need information about the complex object transmission function. Therefore the open-sourced algorithm by Tian et al. \cite{Tian2015} that allows to reconstruct the phase using the formalism of the WOTF from a series of image was implemented in the cellphone's framework. It acquires a series of images while varying the illumination source. \\
To test it's accuracy we mounted a commercially available glass fiber (Thorlabs SMF 28-100, $n(\lambda=530nm)=1.42$) in an immersion medium with $n(\lambda=530nm)=1.46$ and compared it with the expected value of the optical path-difference shown in the graph in Fig \ref{fig6} (a). \\
The processing of the DPC images (Fig \ref{fig6} (e))gives the quantitative phase map shown in (a). The phase jump at the edges of the fiber are due to the discontinuous phase gradient of the object, which is then not satisfying the model of the WOTF anymore. It has to be said, that the optical setup is very robust against misalignments in the optical path such as the position of the cellphone and the condenser relative to each other. A routine helped to automatically "koehler" the setup by centering the illumination aperture by searching for the max. intensity while varying its position on the LCD. Precise quantitative phase reconstructions were possible in almost all cases.\\

\begin{figure}[!htb]
\centering
\includegraphics[width=1\linewidth]{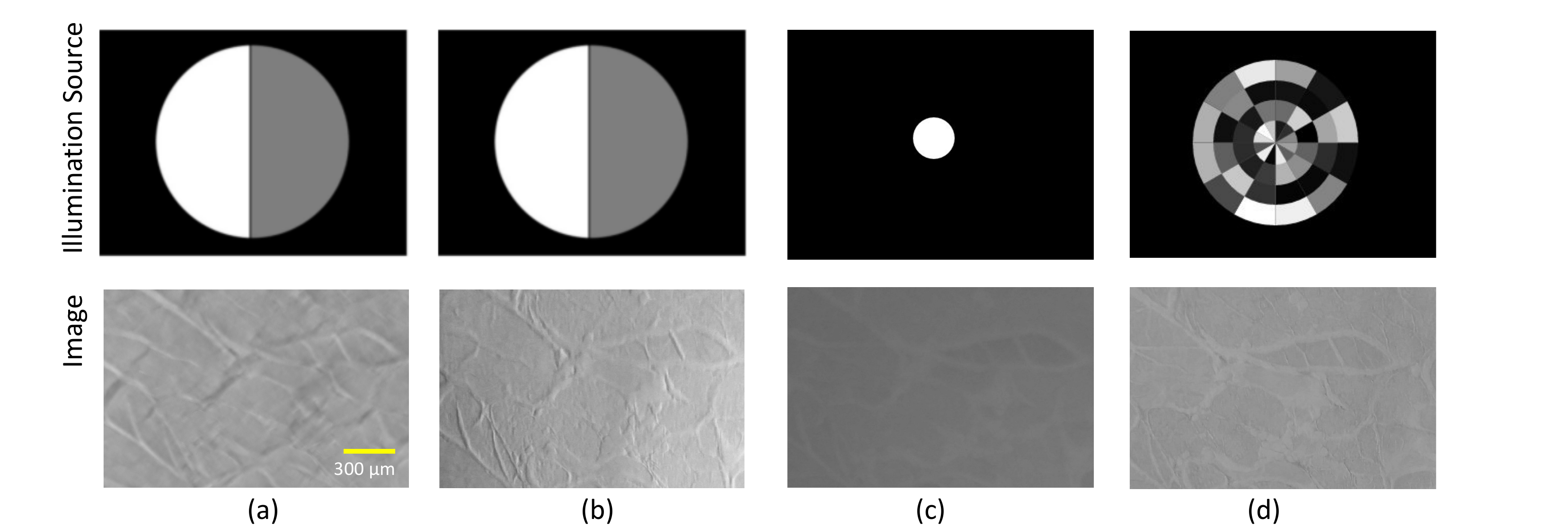}
\caption{\textbf{Quantitative and qualitative results produced by the portable microscope} Quantitatively measured phase of Taste-buds of a rabbit using qDPC mode in (a); the computed DPC image in (b); the intensity measurement and their corresponding illumination sources using brightfield mode with $\na_C=0.2$ (c) and the optimized light-source ($\na_C=0.3$; magnified for better visualization) using the CNN in (h) using (a) as the input image.}
\label{fig7}
\end{figure}

The different acquisition-modes (BF, DPC, qDPC, DF and the optimized illumination-source) are shown in Fig \ref{fig6} (b)-(h), where a series of background and flatfield-images were acquired, prior to the data acquisition, to get rid of possible inhomogeneities in the illumination arrangement given by an incorrect alignment of the projector setup, by simply doing a flatfield-correction. The images are cropped to illustrate smaller details of the object. A non-technical object with a more complicated object structure is shown in Fig \ref{fig7}\\

\subsubsection{Results of the Optimized Light Source}
Giving a quantitative measure of the images quality cannot be defined clearly, because the best contrast remains in the eye of the beholder. Nevertheless we quantized the contrast improvement by computing the error norms which were also used to train the network. Results for the fiber (Fig \ref{S6_Fig}) and the biological sample (Fig \ref{S7_Fig}) are shown in Table \ref{table1} and Table \ref{table2} respectively. The images contrast was calculated by measuring the min/max pixel value and computing $C=\frac{I_{max}-I_{min}}{I_{max}+I_{min}}$, the StDv was calculated using Matlabs $stdfilt$ method. It has to be said, that comparing the two images is degraded by the fact, that the objects, even though very thin, show multiple-scattering effects, which shift i.e. the intensity information along x/y, once illuminated from oblique. \\

\begin{table}[!ht]
\centering
\caption{\bf Contrast measurements of intensity acquisitions of the fiber differently illuminated.}
\begin{tabular}{|l|l|l|l|l|l|}
\hline
\bf Fig. \ref{fig6} & \bf Method and NA & \bf PSNR &	\bf Fidelity &	\bf StDv&	\bf  Contrast\\
\hline
(b) & BF;  $\na_c$=0.1&	53,83	&490,35	&0,64&	0,55\\
(c) & BF;  $\na_c$=0.2&	59,84&	15383,45&	0,42&	0,18\\
(d) & BF;  $\na_c$=0.5&	58,05&	14835,30&	0,51&	0,04\\
(f) & DF;  $\na_c$=0.4&	54,18&	8419,68&	0,65&	0,74\\
(g) & DPC; $\na_c$=0.5&	49,47&	1350,69&	0,68&	0,49\\
(h) & Opt. Illu.;  $\na_c$=0.3&	55,95&	303,77&	0,65&	0,49\\
\hline
\end{tabular}
\begin{flushleft} Comparison of reference and non-reference image quality measurements of the intensity acquisitions done with different illumination shapes from the fiber. The qDPC image was chosen to act as the reference image. 
\end{flushleft}
\label{table1}
\end{table}

\begin{table}[!ht]
\centering
\caption{\bf Contrast measurements of intensity acquisitions of the fiber differently illuminated.}
\begin{tabular}{|l|l|l|l|l|l|}
\hline
\bf Fig. \ref{fig7} & \bf Method and NA & \bf PSNR &	\bf Fidelity &	\bf StDv&	\bf  Contrast\\
\hline
(b) & BF;  $\na_c$=0.2 &46,359	&0,106&	0,0034&	0,27\\
(c) & DPC; $\na_c$=0.5&	50,93&	0,0257&	0,0053&	0,99\\
(d) & Opt. Illu.;  $\na_c$=0.3&	51,63	&0,023&	0,008&	0,39\\
\hline
\end{tabular}
\begin{flushleft} Comparison of reference and non-reference image quality measurements of the intensity acquisitions done with different illumination shapes from the taste bud of a rabbit. The qDPC image was chosen to act as the reference image. 
\end{flushleft}
\label{table2}
\end{table}

These results suggest, that the fidelity (Eq \ref{eq:fidelity}) improves compared to the brightfield mode, while it is not necessary to acquire multiple images and post process them, like in DPC-mode.\\

To demonstrate the effect of the optimized source using the algorithm in Sec \ref{sec:machinelearn} we imaged a histological thin-section of taste-bud cells from a rabbit which show almost no amplitude contrast in BF-mode $\na_c=0.2$ fig. \ref{fig7}  (c). The resulting optimized light source $\na_c=0.3$  is illustrated in Fig \ref{fig7} (d). The dark-field-like aperture, where the frequencies which do not contribute to the image formation are damped by lowering the intensity of the proper segments improve the contrast compared to BF-mode. Introducing the oblique illumination results in a higher coherence factor $S=\frac{\na_{c}}{\na_{o}}>1$, where $\na_o$ gives the numerical aperture from the objective lens. This makes it possible to resolve object structures above the limit given by the typical Rayleigh-criterion $D_{min}=0.61*\frac{\lambda_0}{\na_{eff}}$, where $\na_{eff}= \na_{c}+\na_{o}$ and $D_{min}$ defines the diameter of the smallest resolvable object structure.\\

Resulting from this, the highest possible frequency is given by $k_{max}=\frac{2\cdot NA_{eff}}{\lambda}$, thus the pixel pitch for Nyquist sampling must be less than or equal to $\frac{ NA_{eff}}{2\cdot \lambda}$ which results into a maximum optical resolution of $D_{min}\leq 2.44 \cdot$ pixel size. 
Following the theoretical investigation in \cite{Switz2014}, the theoretical Nyquist-limited resolution of the LG Nexus 5 with a pixel pitch of $1.9\,\mu m$, results into a resolution of $\approx4.4\,\mu m$ at $\lambda=550\,nm$, where the Bayer pattern was already taken into account.\\

Because we used an implicit way to model our optimization procedure using a machine learning approach, one can hardly judge, what the neural network really learned. From the experiments it seems, that the algorithm ``liked'' to give a higher weight to the outer segments of the circular segmented source. This reduces the zeroth order and can eventually increase the contrast because less ``false'' light is used for the image formation as already mentioned earlier. Besides that, it appears, that the method promotes the major diffraction orders of a sample by illuminating the vertically oriented fiber from left/right. This illumination-scheme is also explicitly represented by the so called Archels-Method \cite{Granik2004}.\\

Not all of the quality measures mentioned in Table \ref{table1} are verifying a better contrast of the optimized illumination source, thus it is dare to say, that the effort really improves the images quality. The complexity and large number of parameters of the algorithm can degrade the result of the method. From the experimental results it is clearly visible, that one can see a better phase-contrast, without the use of expensive optics, the drawback of having the phase-gradient like in DPC or only higher order scattering when using DF. The newly introduced method is capable to use the full numerical aperture of the condenser and requires no post-processing of the images once the source-pattern was created.

\section*{Conclusion}

In most cases the computational and imaging potential of a cellphone is not fully exhausted. They serve an integrated framework with an already existing infrastructure for image acquisition and hardware synchronization, as well as for rapid development of user-defined image processing applications. \\

Taking advantage of the widespread availability of high-quality cellphone cameras which are mass-produced in well controlled production process gives a tool at hand which can be further enhanced by adding external hardware, such as the presented automated brightfield-microscope. \\

The field of computational imaging is then able to use the hardware in a way, that new image techniques, such as the qDPC, can be deployed on low-cost devices with high precision. Techniques to surpass the optical resolution limit, e.g. the SMO, that are around for long time can take advantage of new machine-learning applications which gives the opportunity to carry out those, usually computational expensive calculations, on embedded devices as shown in this study. \\
In recent years, the idea of making the sources for hard- and software available to the public, became widely accepted and it enables faster development cycles. Techniques such as the rapid prototyping then offer ways to produce cost effective microscopes e.g. for the use in 3rd-world countries and educational systems.

Further studies e.g. in \cite{Skandarajah2014a, Switz2014, Breslauer2009, Phillips2015} show that quality and characteristic of cellphone cameras improve in each iteration of new hardware. The optical quality of the injection molded cellphone lenses can be used to diffraction limited microscopic imaging with an $NA=0.25$ at price of less then 5\,\$ which opens opportunities to democratize health care or microscopic imaging in general. The results presented here show an optical resolution limit of up to $2\,\mu m$ with the ability to measure the phase quantitatively at high accuracy, as well as imaging adaptive phase contrast methods using an actively controlled illumination source, for a price well below $100\,\$$.\\
The here created platform enables to integrate other imaging techniques, such as the Fourier Ptychography (FPM) \cite{Tian2014a, Ou2015, Zheng} or include further image processing in the cellphone's software such as tracking and classification of biological cells \cite{Angermueller2016}. 



%

\section*{Acknowledgments}
The authors thank Joerg Sprenger and Marzena Franek (CZ Microscopy GmbH) for helpful conversations as well as support and Jörg Schaffer (CZ Microscopy GmbH) for providing the research microscope. 

\nolinenumbers

{\footnotesize
    \bibliography{./library}}
\bibliographystyle{alphadin} 

\end{document}